\documentclass[journal]{IEEEtran}
\usepackage{amsmath,amsfonts}
\usepackage[ruled,noline,longend,nofillcomment]{algorithm2e}
\usepackage{array}
\usepackage{subcaption}
\usepackage{textcomp}
\usepackage{stfloats}
\usepackage{url}
\usepackage{verbatim}
\usepackage{graphicx,cite,booktabs,multirow,makecell}
\usepackage{hyperref}
\usepackage{orcidlink}
\hypersetup{
	colorlinks=true,
	linkcolor=blue,  % 内部链接颜色为黑色
	citecolor=blue,  % 引用链接颜色为黑色
	urlcolor=magenta,   % URL 链接颜色为黑色
	linkbordercolor=0 0 0, % 边框颜色设置为透明，即不显示边框
}
\usepackage[numbers,sort&compress]{natbib}
\begin{document}

\title{Inductive Graph Neural Networks for Node Centrality Approximation in Complex Networks}
\author{Yiwei Zou$^{\orcidlink{0009-0002-9288-8280}}$, Ting Li$^{\orcidlink{0009-0009-2105-5602}}$, and Zong-fu Luo$^{\orcidlink{0000-0003-4871-4216}}$ \thanks{This work was supported by the Fundamental Research Funds for the Central Universities, Sun Yat-sen University (No. 23QNPY78) and the National Laboratory of Space Intelligent Control (No. HTKJ2023KL502003). (Corresponding author: Zong-fu Luo.)} \thanks{The authors are with the School of Systems Science and Engineering, Sun Yat-sen University, Guangzhou 510275, People's Republic of China (E-mail: zouyw3@mail2.sysu.edu.cn; liting226@mail2.sysu.edu.cn; luozf@mail.sysu.edu.cn).}}	

%\markboth{Journal of \LaTeX\ Class Files,~Vol.~14, No.~8, August~2023}%
%\IEEEpubid{0000--0000/00\$00.00~\copyright~2023 IEEE}
\maketitle
\begin{abstract}
Closeness Centrality (CC) and Betweenness Centrality (BC) are crucial metrics in network analysis, providing essential reference for discerning the significance of nodes within complex networks. These measures find wide applications in critical tasks, such as community detection and network dismantling. However, their practical implementation on extensive networks remains computationally demanding due to their high time complexity. To mitigate these computational challenges, numerous approximation algorithms have been developed to expedite the computation of CC and BC. Nevertheless, even these approximations still necessitate substantial processing time when applied to large-scale networks. Furthermore, their output proves sensitive to even minor perturbations within the network structure.

In this work, We redefine the CC and BC node ranking problem as a machine learning problem and propose the CNCA-IGE model, which is an encoder-decoder model based on inductive graph neural networks designed to rank nodes based on specified CC or BC metrics. We incorporate the MLP-Mixer model as the decoder in the BC ranking prediction task to enhance the model's robustness and capacity. Our approach is evaluated on diverse synthetic and real-world networks of varying scales, and the experimental results demonstrate that the CNCA-IGE model outperforms state-of-the-art baseline models, significantly reducing execution time while improving performance.
\end{abstract}


\section{Introduction}
\IEEEPARstart{N}{etwork} models are widely used in several real-world scientific fields, including complex networks \cite{ws, ba}, computer science \cite{1}, biology \cite{2} and sociology \cite{3,67}. Research has shown that a few specific nodes within the network can significantly affect a network's performance. The malfunction or activation of these nodes, commonly known as critical nodes, can dramatically impact various network functions \cite{4}. Node centrality metrics for networks are crucial for analyzing networks and identifying key nodes based on their relative importance. However, the computation of centrality metrics becomes complex and time-consuming when applied to real-world complex networks with thousands or even millions of interconnected nodes and edges \cite{5}.

Common node centrality metrics include degree centrality, betweenness  centrality  and  closeness  centrality,  etc. These metrics, which define the concept of centrality of a node from different perspectives, are used to identify key nodes in complex networks. The computational complexity of different centrality metrics varies significantly according to the calculation formula. In general, degree centrality is considered to have lower computational complexity, while  betweenness  centrality  and  closeness  centrality  bear higher complexity. However, the latter two metrics find broad applications in community  detection \cite{CD} and  network disassembly \cite{ND}. Computing the high-complexity centrality of all nodes directly might be impractical, since many real network models are large-scale and complex. Echoing this requirement, how to combine existing machine learning and neural network methods to approximately compute high-complexity centrality metrics with low-complexity centrality metrics by their correlations has become a hot topic \cite{6,7,8}. 

% TODO
In this paper, we propose a Complex Network Centrality Approximation using Inductive Graph Embedding (CNCA-IGE) model. The degree centrality metrics of each node in the network are used as node features, firstly, the inductive graph embedding methods Graph SAmple  and  aggreGatE  (GraphSAGE) \cite{23} and  Variational Graph AutoEncoder (VGAE) \cite{21} are used to map the nodes in  the  network  into  embedding  vector  representations,  and secondly, the embedding vectors are used as inputs, and the Multilayer Perceptron (MLP) \cite{35} and Multilayer Perceptron Mixer (MLP-Mixer) neural network architectures are chosen to train the regression model. The regression model is  able to approximate the high computational complexity closeness centrality ranking and betweenness centrality ranking with low computational complexity degree centrality. The model parameters are trained in an end-to-end manner, where the training data consists of synthetic networks. We conducted extensive experiments on synthetic networks represented by small-world and scale-free networks and on six real-world complex networks with various sizes. To the best of our knowledge, our method outperforms state-of-the-art baseline models for centrality ranking on synthetic and real-world networks. In terms of running time, our model is more efficient than the one based on transductive node embedding. Meanwhile, experiments show that the regression model achieved through training with the MLP-Mixer decoder is significantly superior in terms of robustness and generalisation.

The main contributions of this paper are summarized as follows:

(1)  We transform the CC and BC ranking problem of nodes into a machine learning problem. Propose an inductive graph neural network-based encoder-decoder model, CNCA-IGE, for ranking nodes in a network based on specified CC or BC metrics.

(2)  We propose to use the MLP-Mixer model as decoder in the BC ranking prediction task. Its added internal feature mixing of node embedding vectors facilitates the enhancement of model capacity. In addition, the neural network architectural components such as residual connectivity and layer normalisation employed by MLP-Mixer help to build more robust prediction models.

The paper is structured as described below. In Section \ref{sec2}, we summarise the mainstream graph embedding methods and investigate the research in progress of network centrality prediction. Section \ref{sec3} outlines the centrality metric ranking prediction model CNCA-IGE introduced in this paper and the associated training procedure. Specific experimental results and discussions are presented in Section \ref{sec4}, and Section \ref{sec5} concludes the paper.

\section{Related Work}\label{sec2}
\subsection{Unsupervised Graph Embedding}
Graph embedding maps nodes or edges in a graph to a low-dimensional vector representation through a set of weight matrices. By utilizing the adjacency and feature matrices as inputs, embedding vectors are generated to reflect the graph's topological and connectivity characteristics of nodes or edges. In the following, we explore three mainstream models for generating graph embeddings.

\subsubsection{Matrix Factorization}~\par

Early techniques for generating vector representations on graphs involved Matrix Factorization. Laplacian Eigenmaps (LE) \cite{9} is a groundbreaking algorithm that utilizes this method. LE aims to project nodes with vital first-order proximity into similar vectors in the embedding space. To embed the graph in $d$ dimensions, LE constructs a similarity matrix, computes the Laplacian of the graph, and then determines its eigenvectors corresponding to the $d$ smallest eigenvalues.

\subsubsection{Random Walk techniques}~\par
The authors of Deepwalk \cite{10} utilized a word embedding architecture similar to Word2Vec \cite{11} to generate vector representations for the nodes in a graph. During the data preparation phase, random walks are conducted on the graph to generate sequences of nodes equivalent to the sentences in Word2Vec. Then, a sliding window moves across each sequence to develop segments of consecutive nodes that will be used as training data for the model. During the training phase, a SkipGram model is used on the data produced in the first section. The model aims to predict the neighbors of each middle node in the sliding window. DeepWalk strives to map nodes with similar neighborhoods into comparable vectors in the embedding space.

DeepWalk randomly selects nodes during the walk, while Node2Vec \cite{12} introduces parameters $p$ and $q$ to control the sampling. The parameter $p$ influences the likelihood of returning to a node $v$ after visiting another node $t$, and the parameter $q$ affects the likelihood of departing from node $v$ once it has been visited.

\subsubsection{Deep Neural Network}~\par
Progress in deep learning has resulted in a new area of research focused on utilizing neural networks for graph data \cite{14,15}. Structural Deep Network Embedding (SDNE) \cite{16} and Deep Neural networks for learning Graph Representations (DNGR) \cite{17} utilize deep autoencoder to capture non-linearity in graphs and, at the same time, perform dimension reduction for generating graph embedding. Graph Convolutional Network (GCN) \cite{18} provides a simplified approximation to spectral convolution and improves computational efficiency for semi-supervised multi-class node classification, making it suitable for various machine learning tasks. Enhancements have been suggested to boost the training speed of GCNs in later studies \cite{19,20}.

Lately, researchers have suggested utilizing deep autoencoders to acquire compressed representations that capture the essence of graph structure. An autoencoder comprises an encoder and a decoder collaborating to minimize reconstruction loss. A model designed specifically for graphs is the Graph AutoEncoder (GAE), which features a Graph Convolutional Network (GCN) encoder for generating embeddings and an inner product decoder to reconstruct the adjacency matrix ($\hat{A}=\sigma(UU^T)$). VGAE is a probabilistic version of GAE. It introduces a distribution over latent variables $Z$, with these variables being conditionally independent Gaussians given $A$ and $X$ with means ($\mu$) and diagonal covariances ($\sigma$) being parameterized by two GCN encoders \cite{22}. As in the case of images, VGAE just adds KL-divergence term between conditional distribution $q(Z|X,A)$ and unconditional $p({Z})\sim N(0,1)$ to the loss. After node embeddings are reconstructed via random normal distribution sampling, that is, $Z=\mu+\sigma\varepsilon.$. Then adjacency matrix is decoded using inner product of achieved vector $Z$ as in simple GAE.

In a recent study, the authors of GraphSAGE present an expansion of GCN for inductive unsupervised representation learning and propose using trainable aggregation functions rather than simple convolutions applied to neighborhoods in GCN. GraphSAGE learns how to aggregate information from various neighborhood depths to create node representations based on their initial features. To better illustrate the complex connections among nodes in more detail, GAT \cite{24} utilizes masked self-attention layers to learn weights that balance the influence of neighbors on node embedding, accommodating both inductive and transductive learning scenarios. Similar to GCN, GAT contains several hidden layers $H^i=f(H^{i-1},A)$, where $H_0$ is a graph node features. In each hidden layer linear transformation of input is firstly calculated with the learnable matrix $W$. The authors have substituted the adjacency matrix with a learnable self-attention mechanism in the form of a fully-connected layer. This layer includes activation functions and normalization with softmax.

\subsection{Network Centrality Prediction}
The computation of node centrality measures is of great significance to the study of complex networks. It is often used to evaluate or identify the importance of different nodes in the network. However, in practice, the computational complexity of different centrality measures varies greatly. Some commonly used centrality measures, such as closeness centrality and betweenness centrality, even with the latest advanced optimization algorithms (Brande's algorithm \cite{25} and its variants), have $O(mn)$ complexity, which limits the application of centrality analysis to complex networks. In order to apply the centrality measures to large-scale real networks, there are two main improvements: one is to use a distributed computational approach to extend the centrality computation from a single machine to a cluster of high-performance computers. \cite{26,27,28}. The other is to use approximation computation to sacrifice accuracy for higher computational efficiency. Early proposed schemes are the sampling-based betweenness centrality and closeness centrality approximation methods \cite{29}. which determine the centrality metrics of a node by computing the single-source shortest paths (SSSP) of a specific sample of nodes and then estimating the centrality metrics of other non-sampled nodes using SSSP. Moreover, improvements in previous methods have been proposed by adding guaranteed value of error \cite{30} and adaptive evolutionary graph sampling techniques \cite{31}. Despite the advancements in approximating node centrality measures, the sampling techniques still expensive in computation due to the high complexity of calculating precise centrality values for a small fraction of nodes in complex networks.

With the emergence of technologies in the field of artificial intelligence such as machine learning and neural networks, more scholars have focused on training neural network models to approximate network centrality measures. Recently, Grando et al. \cite{7,8} proposed a regression model based on a multilayer perceptron that trains on the adjacency matrix of a graph and two sets of centrality measures for each node (degree centrality and eigenvector centrality) as inputs. This model is used to predict the remaining centrality measures of the nodes. Chen et al. \cite{32} further improved Grando's model by using a pointwise learning-to-rank algorithm to transform the regression problem of predicting centrality values into a pairwise ranking problem. They trained a neural network-based ranking model to predict the closeness centrality ranking of nodes. 

However, using only neural network models cannot effectively capture and utilize the features and topological information of nodes and their neighbors in a graph. Graph Neural Networks (GNNs) can be used to aggregate node features and generate representation vectors (typically used for dimensionality reduction), which helps fully extract node features and reduce the model's parameters. Recently, there have been studies based on the idea of Deep Learning (DL) that design an Encoder-Decoder architecture. In this architecture, the Encoder adopts GNN algorithms to map the adjacency matrix A and feature vectors $X$ (typically the computationally efficient degree centrality d) of the graph to low-dimensional embedding vectors H. The Decoder then transforms the embedding vectors H into output vectors $Y$ of node centrality scores or rankings. Fan et al. \cite{33} designed a GNN model based on Gated Recurrent Unit (GRU), where the upstream encoder aggregates node neighbor features using neighbor sampling and weighted summation, and uses GRU to decide which parts of the information to retain. After multiple iterations, the maximum feature vector is selected as the node's embedding vector. The downstream decoder uses two MLP layers to map the embedding vector to betweenness centrality ranking scores.  Maurya et al.  designed a variant of GNNs specifically for CC and BC. They modified the adjacency matrix through preprocessing to restrict the paths of feature propagation. Afterward, a rank-based loss was utilized for training a scoring function. This method computes the out-degree and in-degree characteristics individually, allowing it to be used with directed graphs. 

The beforementioned DL methods manually constructed the GNNs or loss functions based on the properties and computational principles of the target centrality measures. While they perform well in predicting specific centrality measures, they are not conducive to extending to predicting remaining centrality measures with different computational principles. Currently, there have been studies on general frameworks that can use a single model to predict different types of centrality measures. Mendonça et al. \cite{36} combined transductive graph embedding techniques, such as GCN and Struc2Vec, with regression models based on Grando's work. The regression models were used to predict the values or rankings of any target centrality measure, and the Mean Squared Error (MSE) was computed as the loss function by comparing the predictions with the ground truth. However, transductive graph embedding methods have limitations in terms of generalization from the training set to real-world network datasets \cite{23}. Additionally, due to the large scale and dynamic changes in the topology of real-world complex networks caused by the addition and removal of nodes, transductive graph embedding methods require retraining the model every time the graph topology changes, leading to high computational and storage costs.

\section{Methodology}\label{sec3}
In this section, we introduce the proposed method for approximating the closeness ranking and betweenness ranking in large-scale real-world networks. We start with the preliminaries about input feature selection, followed by a description of the model implementation details and the training algorithm, with the complexity analysis to be developed at the end.

\subsection{Preliminaries}
\subsubsection{Centrality Metric}~\par
\textbf{Degree centrality (DC)} represents the most simple centrality metirc. The Degree centrality of node $i$ is given by:
\begin{equation}
d(w)=\sum_{j\in V} a_{ij}
\end{equation}
where $a_{ij}$ denotes the element in row $i$ and column $j$ of the adjacency matrix. However, this centrality is inadequate to describe some node features, prompting the construction of a more relevant centrality measure.

\begin{figure*}[htpb]
	\centering
	\includegraphics[width=0.6\linewidth]{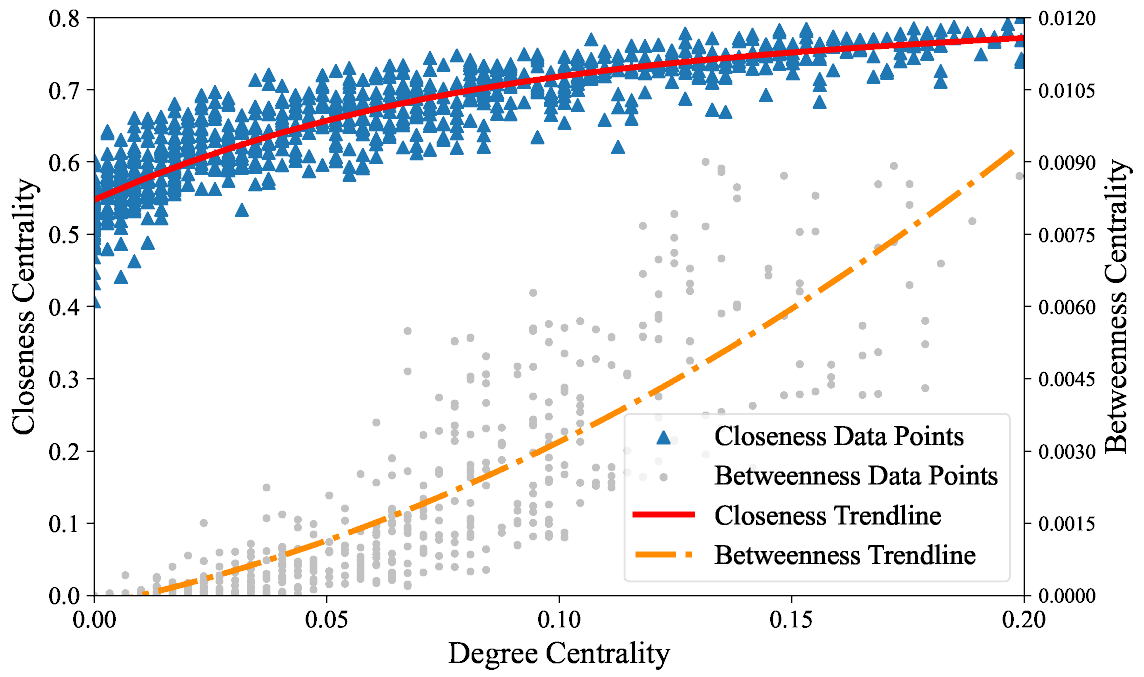}
	\caption{Degree Centrality's Correlation with Closeness and Betweenness Centrality}
	\label{fig:01}
\end{figure*}

\textbf{Closeness centrality (CC)} for node $i$ is defined as the inverse of the mean minimum distance from that node to all other $N - 1$ nodes in the network, as provided by
\begin{equation}
c\left(i\right)=\frac{N-1}{\sum_{j\neq i\in V}{\delta(i,j)}}
\end{equation}
where $\delta(i,j)$ is the distance between node $i$ and $j$.

\textbf{Betweenness centrality (BC)} measures the significance of particular nodes based on the proportion of shortest routes that pass through them. Formally, the normalized BC value $b(w)$ of a node $w$ is defined:
\begin{equation}
b\left(w\right)=\frac{1}{|V|(\left|V\right|-1)}\sum_{u\neq w\neq v}\frac{\sigma_{uv}(w)}{\sigma_{uv}}
\end{equation}
where $|V|$ represents the number of nodes in the network, $\sigma_{uv}$ denotes the number of shortest paths from $u$ to $v$, $\sigma_{uv}(w)$ denotes the number of shortest paths from $u$ to $v$ that pass through $w$.

\subsubsection{Assumption}~\par
The essential premise of machine learning-based proximation approaches is that high-complexity centrality measures may be effectively approximated by feeding low-complexity centrality measurements into the model. Because low-complexity centrality measures are chosen based on their theoretical foundations and relationship to betweenness and closeness centrality. We chose the email-Eu-core network as an example network\footnote{ Download from \url{http://snap.stanford.edu/data/} [Valid until Sep. 2023]}. The network consists of 1005 nodes and 25571 edges, representing the members of a large European research organization and the email correspondence that exists between the members, respectively. In parallel, the correlation between degree centrality and closeness as well as betweenness centrality is shown in Fig. \ref{fig:01}. It can be observed intuitively that nodes with higher degree centrality also have relatively higher closeness centrality and betweenness centrality. Most of the real-world networks in our experiments show the same correlation.
% \begin{figure}[htbp]
% 	\centering
% 	\begin{subfigure}{0.49\linewidth}
% 		\centering
% 		\includegraphics[width=0.99\linewidth]{image/correlation_closeness.pdf}
% 	\caption{Degree/Closeness}
% 	\end{subfigure}
% 	\centering
% 	\begin{subfigure}{0.49\linewidth}
% 		\centering
% 		\includegraphics[width=0.99\linewidth]{image/correlation_betweenness.pdf}
% 	\caption{Degree/Betweenness}
% 	\end{subfigure}
% 	\caption{Degree Centrality's Correlation with Closeness and Betweenness Centrality}
% 	\label{fig:01}
% \end{figure}

% \begin{figure}[htpb]
% 	\centering
% 	\includegraphics[width=0.95\linewidth]{image/correlation.pdf}
% 	\caption{Degree Centrality's Correlation with Closeness and Betweenness Centrality}
% 	\label{fig:01}
% \end{figure}

\textcircled{1}	Degree Centrality vs Clossness Centrality

The essential principle behind degree centrality is that the center node is linked to as many nodes as feasible. Theoretically, closeness centrality is positively associated with degree centrality since a node that can be reached by many other nodes in fewer steps must be well-connected.

\textcircled{2}	Degree Centrality vs Betweenness Centrality

Node betweenness centrality is often associated with its degree centrality, as high betweenness nodes function as critical bridges in the network by linking numerous shortest paths between nodes. However, not all highly connected nodes are essential mediators. Consequently, although node betweenness centrality generally increases with its degree centrality, this relationship doesn't universally apply, indicating a more non-linear correlation between the two factors.

On the whole, the degree centrality is positively correlated with the closeness centrality and betweenness centrality. Therefore, the degree centrality is chosen as an input feature for our model.

\subsection{Model Structure}
\begin{figure*}[!t]
 	\centering
 	\includegraphics[width=0.95\linewidth]{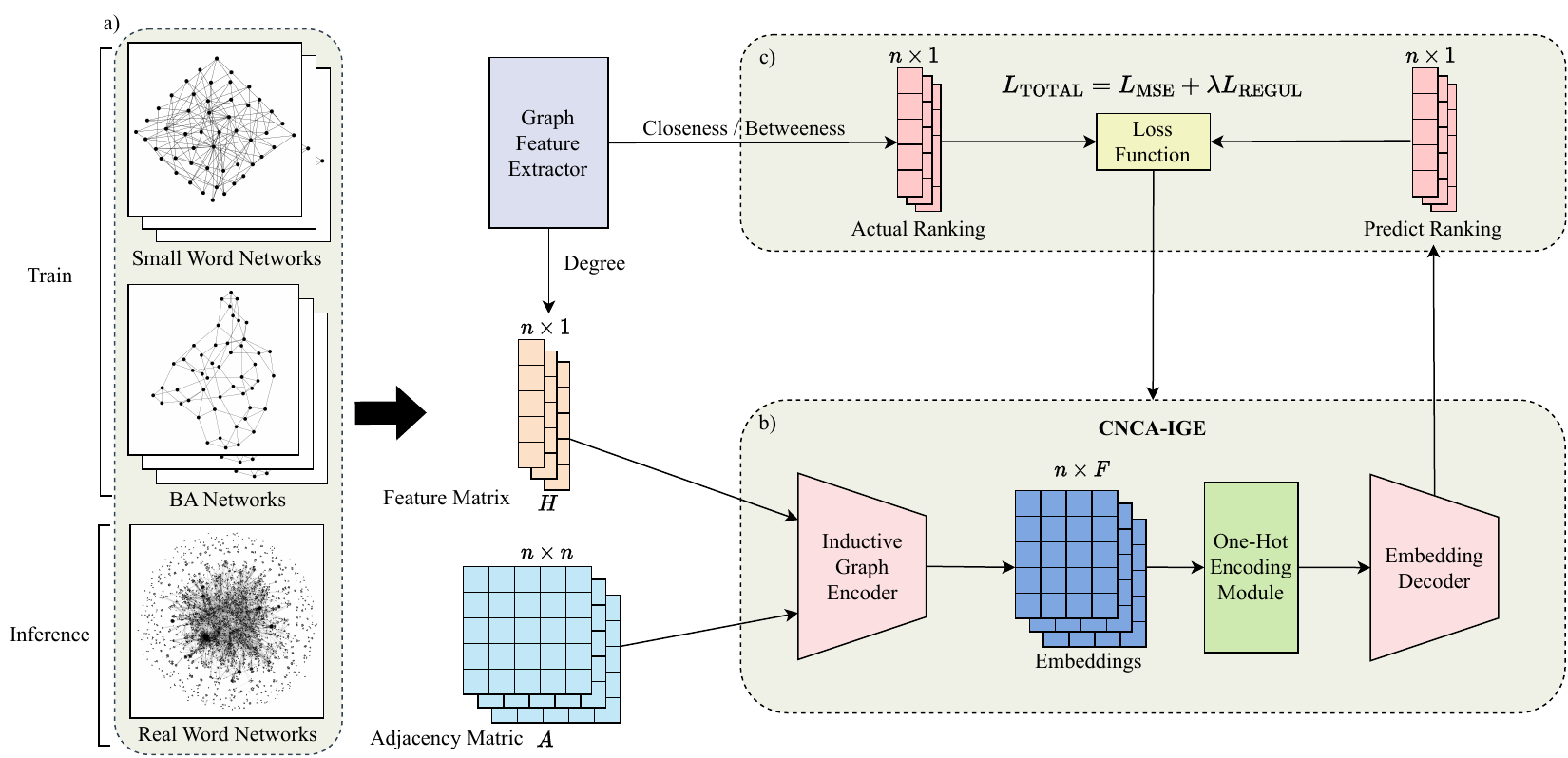}
 	\caption{Overview of CNCA-IGE Model. \textbf{(a)} Dataset: Synthetic "Small-World Networks" and "Scale-Free Networks" are employed as training datasets, and "real-world complex networks" are employed as validation datasets. \textbf{(b)} Model Pipeline: Taking the adjacency matrix and the feature matrix (degree) as inductive graph embedding inputs, the embedding vector is obtained after the One-Hot Encoding module, and then the node ranking is obtained through the decoder module. \textbf{(c)} Training Stage: The loss function is constructed based on the actual node rankings and the predicted node rankings to train the model.}
 	\label{fig:03}
 \end{figure*}

The overall model structure is shown in Fig. \ref{fig:03}. It consists of two parts: the inductive graph embedding and the neural network.

Inductive Graph Embedding: Here, we leverage the degree centrality as a node feature. We utilize the adjacency matrix of the graph and the feature vectors of nodes as input to one of two inductive graph embedding methods, VGAE or GraphSAGE. The output of this graph embedding module provides a low-dimensional vector representation for each node in the network. Notably, our inductive approach differs from transductive graph embedding, as it focuses on constructing a prediction model. Consequently, it eliminates the need to re-run the algorithm for training when new data nodes are encountered.

Neural Networks: In this part of our model, we utilize the embedded matrix $H$, which is generated by the upstream graph embedding module, as input. We employ either an MLP  or MLP-Mixer to train a regression model. This regression model has the capability to predict the ranking of both closeness centrality and betweenness centrality for networks of varying sizes. It is worth noticing that MLP-Mixer offers an advantage over traditional MLP by considering the internal feature mixing of the embedded node vectors, resulting in a larger model capacity and the ability to learn more intricate and complex patterns.

\subsection{CNCA-IGE}
\subsubsection{Inductive Graph Embedding}

%\subsection*{{\normalsize \textbf{GraphSAGE}}}
~\par
\begin{figure*}[htpb]
	\centering
	\includegraphics[width=0.8\linewidth]{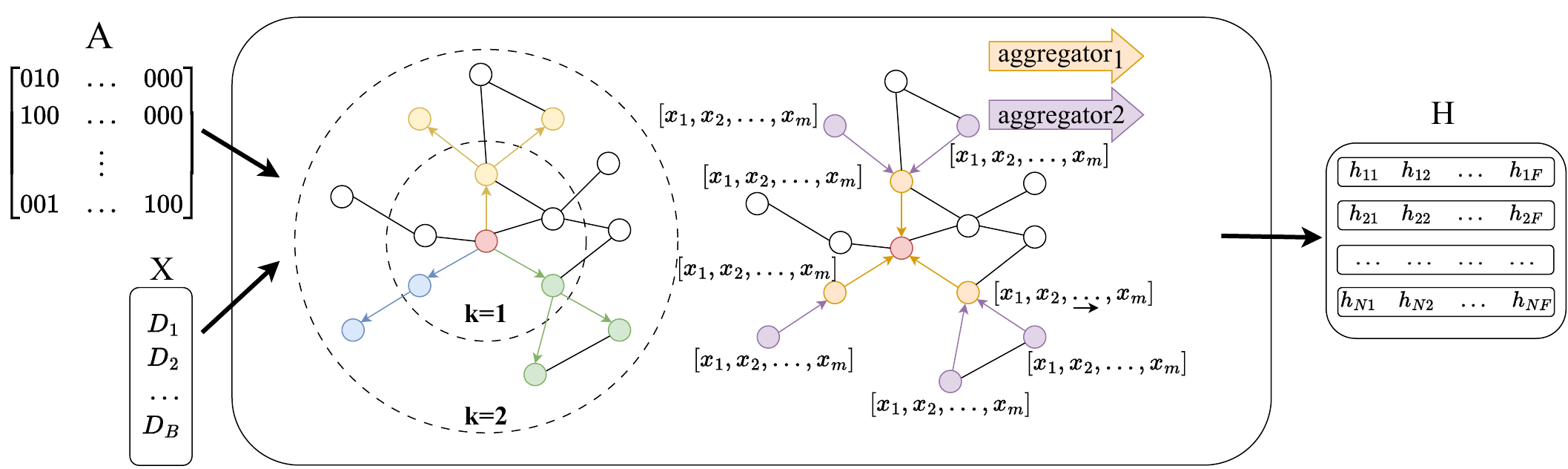}
	\caption{Architectural Overview of GraphSAGE}
	\label{fig:04}
\end{figure*}

\textbf{GraphSAGE} is a technique that extracts feature information by leveraging the feature vectors $X$ of nodes and the adjacency matrix $A$ of the graph. As shown in the Fig. \ref{fig:04}, the process involves multiple rounds of iterations, with each iteration updating a node's feature vector by aggregating information from its neighboring nodes. To efficiently handle nodes in large-scale networks, GraphSAGE employs neighbor sampling, where only a subset of neighbor nodes is selected for feature aggregation. 

Following neighbor sampling, an aggregator is utilized to combine the feature vectors of the selected neighbor nodes, resulting in an embedding vector representation for the target node. This aggregator can take the form of a simple averaging or pooling operation, or it may involve more intricate mechanisms such as attention mechanisms or graph convolution operations.

The pooling aggregator is both symmetric and trainable. The pooling aggregator sequentially performs nonlinear transformation, pooling operation on the neighborhood vectors. Then the obtained result is concatenated to the current node's vector and performs another nonlinear transformation in order to get the updated embedding vectors of the node. The formula of pooling aggregator is as follows:
\begin{equation}\label{4}
\text{AGGREGATE}_{k}^{\text{pool}} = \max \left( \left\{ \sigma \left( W_{\text{pool}}^k h_{u_i} + b \right)  \right\} \right)
\end{equation}
$\forall u_i \in N(v)$, where $\sigma$ denotes nonlinear activation function, $W_{pool}$ denotes a set of learnable weight matrices, and $h_{u_i}^k$ denotes the neighborhood embedding vector representation of the node $V$. Through sampling and aggregation, GraphSAGE is able to learn the embedding matrix $H$ of all nodes in the graph.

\begin{figure*}[htpb]
	\centering
	\includegraphics[width=0.9\linewidth]{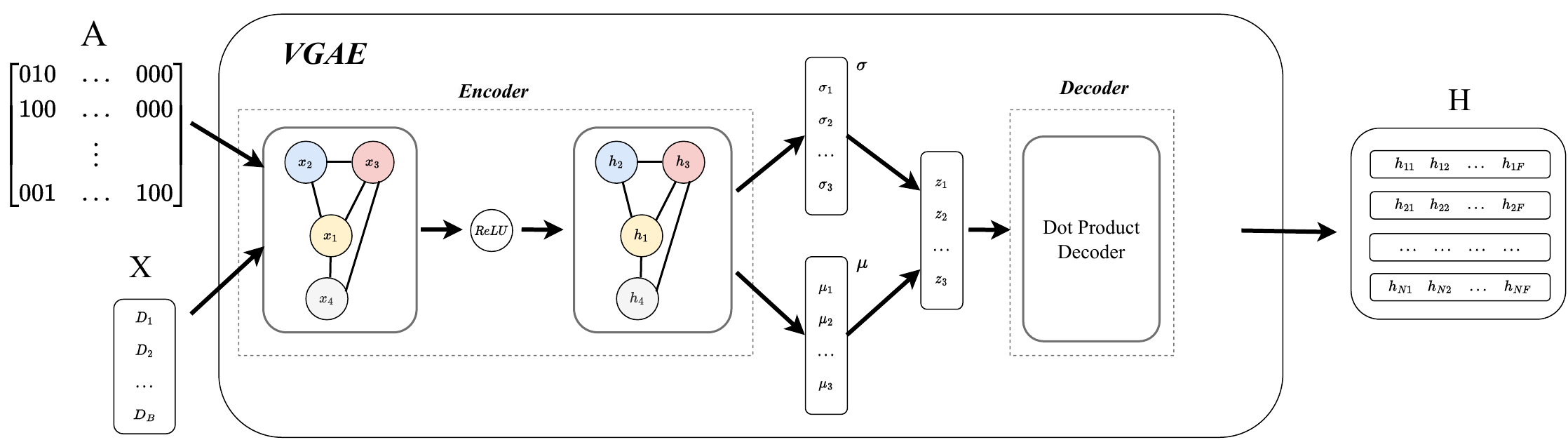}
	\caption{Architectural Overview of VGAE}
	\label{fig:05}
\end{figure*}

%\subsection*{{\normalsize \textbf{VGAE}}}
\textbf{VGAE} presents an inductive framework by merging auto-decoding and variational inference. As shown in the Fig. \ref{fig:05}, VGAE takes the graph's adjacency matrix $A$ and node feature matrix $X$ as its input. The core idea of VGAE is to utilize the graph structure to learn the mean $\mu$ and variance $\sigma$ of low-dimensional node vector representations through an encoder. These learned parameters define the distribution of the node vector representations. The final embedded vector representation is derived by sampling from this distribution.

The encoder consists of a two-layer GCN:
\begin{equation}\label{5}
\begin{aligned}
q\left(z_i\ \right|\ X,\ A)&=N\left(z_i\ \right|\ \mu_i,\ diag(\sigma_i^2))	\\
		q\left(Z\ \right|X,A)&=\ \prod_{i=1}^{N}{q\left(z_i\ \right|\ X,A)}
\end{aligned}
\end{equation}

$\mu$ is the mean of the node vector representation ($\mu={\text{GCN}}_\mu(X,A)$), $\sigma$ is the variance of the node vector representation ($\log\sigma={\text{GCN}}_\sigma(X,A)$). Note that ${\text{GCN}}_\mu(X,A)$ and ${\text{GCN}}_\sigma(X,A)$ share $W_0$ but not $W_1$, and the sampling variables use the reparameterization trick to avoid the inability to perform gradient backpropagation due to the objective function being non-differentiable as a result of sampling.

The decoder reconstructs the graph network by computing the probability of the existence of an edge between two nodes in the graph topology:
\begin{equation}
p\left(A\ \right|\ Z)=\prod_{i=1}^{N}\prod_{j=1}^{N}{p\left(A_{ij}\ \right|\ Z_i,Z_j)}
\end{equation}
where $p\left(A_{ij}=1\right|\ Z_i,Z_j)={\rm sigmoid}(z_i^Tz_j)$.

The loss function consists of two parts: 
\begin{equation}
L=E_{q\left(Z\ \right|X,\ A)}[\log p\left(A\ \right|\ Z)]-KL[qZ  X,A)||p(Z)]
\end{equation}
where $E_{q\left(Z\right|X,A)}[\log p\left(A\right|Z)]$ is the distance measure between the reconstructed and original graphs, and $KL[q\left(Z \right|X,$ $ A)|\left|p\left(Z\right)\right]$ is the Kullback-Leibler divergence between $q(\cdot)$ and $p(\cdot)$.

VGAE measures the difference between the reconstructed graph and the original graph by applying random noise to the node embedding vectors generated by the encoder. By minimizing the reconstruction loss, VGAE learns the embedding matrix $H$ $(N\times F)$ of all nodes in the graph.


\subsubsection{Neural Network}~\\
$~$\quad After obtaining the embedding matrix $H$ $(H\in\mathbb{R}^{N\times F})$, we work on the embedding representation of each node in the graph in batches. By multiplying the one-hot coding matrix $I_D$ of each batch of nodes with the embedding matrix $H$, we are able to unify the dimensions of the matrix $H_D$ as input to the downstream neural network:
\begin{equation}
H_D=I_D\times H
\end{equation}
where $I_D\in\mathbb{R}^{B\times N}$, $H_D\in\mathbb{R}^{B\times F}$, $N$ denotes the number of nodes in each graph, $B$ denotes the number of nodes in each batch, and $F$ is the dimension of the embedding vectors from upstream graph embedding module.

The downstream prediction model can choose either MLP or MLP-Mixer, and both sets of downstream models take $H_D$ as input and output the predicted centrality ranking $Y$.

%\subsection*{{\normalsize \textbf{MLP}}}
\textbf{MLP} prediction model consists of three hidden layers and one output layer with the following architecture: 
\begin{equation}\label{9}
Y = {\rm ReLU}({\rm ReLU}({\rm ReLU}(H_{{D}}W^{F1})W^{F2})W^{F3})W^{F4}
\end{equation}
where $W^{F1},W^{F2},W^{F3},W^{F4}$ are all weight matrices.

\begin{figure*}[htbp]
	\centering
	\begin{subfigure}{0.675\linewidth}
		\centering
		\includegraphics[width=0.9\linewidth]{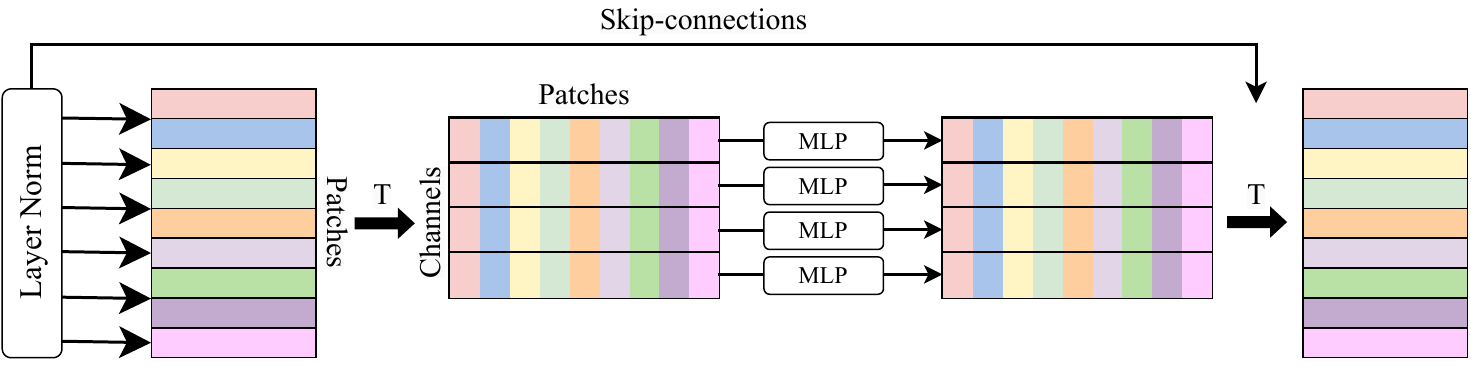}
	\caption{Token-Mixing Module}
	\end{subfigure}%
	\centering
	\begin{subfigure}{0.325\linewidth}
		\centering
		\includegraphics[width=0.9\linewidth]{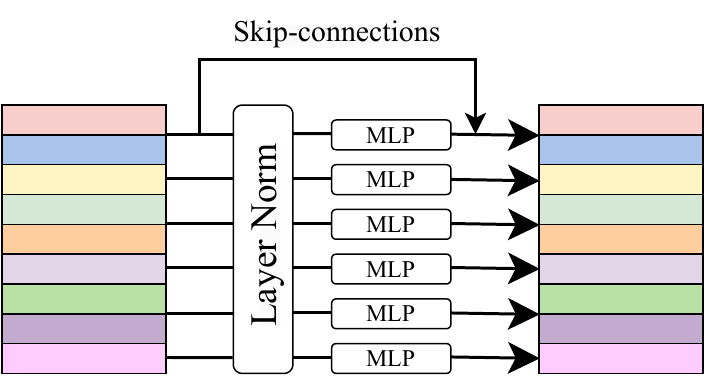}
	\caption{Channel-Mixing Module}
	\end{subfigure}
	\caption{Architectural Overview of MLP-Mixer}
	\label{fig:06}
\end{figure*}

%\subsection*{{\normalsize \textbf{MLP-Mixer}}}
\textbf{MLP-Mixer} prediction model consists of two Mixer modules of the same size with the following architecture:

% \begin{figure*}[htpb]
% 	\centering
% 	\includegraphics[width=0.8\linewidth]{image/06}
% 	\caption{MLP-Mixer Architecture}
% 	\label{fig:06}
% \end{figure*}

As shown in the Fig. \ref{fig:06}, the first Mixer module is the token-mixing module, which acts on the columns of $H_D$ (i.e., it applies to ${H_D}^T$), the mapping space is $\mathbb{R}^B\rightarrow\mathbb{R}^B$, and all MLP layers share the same parameters. The second Mixer module is the channel-mixing module, which applies to the rows of $H_D$, the mapping space is $\mathbb{R}^F\rightarrow\mathbb{R}^F$, and likewise, all MLP layers share the same parameters. Each MLP module contains two fully connected layers and a nonlinear activation function (here the GELU function is used) applied independently to each row of its input data tensor. In addition to the MLP layers, MLP-Mixer uses neural network architecture components such as residual connection and layer normalization. Eventually, the result of Mixer is output by the hidden layers and final output layer.

The downstream prediction model of MLP-Mixer is structured as follows:
\begin{equation}\label{10}
\begin{aligned}
U_{\ast,i}&=H_{D_{\ast,i}}+\mathbf{W}_2\sigma(\mathbf{W}_1\mathrm{\ LayerNorm\ }(H_D)_{\ast,i}),\\
&\qquad\qquad\qquad\qquad\qquad\qquad\quad\mathrm{\ for\ }i=1\ldots B\\
Y_{j,\ast}&=U_{j,\ast}+\mathbf{W}_2\sigma(\mathbf{W}_1\mathrm{\ LayerNorm\ }(U)_{j,\ast}),\\
&\qquad\qquad\qquad\qquad\qquad\qquad\quad\mathrm{\ for\ }j=1\ldots F
\end{aligned}
\end{equation}
where $\sigma$ is the Gaussian error linear unit activation function.

Since the size of the input matrix $H_D\in\mathbb{R}^{B\times F}$ is fixed for different graph sizes $N$ (only with respect to the batch size $B$ and the dimension $F$ of the output embedding vectors), the model is capable of predicting node centralities for graphs of any size.

\subsection{Training Algorithm}
Algorithm \ref{alg1} describes the training algorithm for CNCA-IGE. For the MLP and MLP-Mixer neural network models, the error function we choose is the MSE between the predicted centralities and actual values. Moreover, L2 regularization function is added to avoid overfitting. Hence the total loss function is:
\begin{equation}\label{11}
L_{\text{TOTAL}} = L_{\text{MSE}} + \lambda L_{\text{REGUL}}
\end{equation}
where $L_{\text{MSE}}$ is the loss function of the mean square error, $L_{\text{REGUL}}$ is the L2 regularization loss, and $\lambda$ is the weight of controling the regularization term. We use the Adam optimizer as the optimization function and the gradient is clipped to the range $[-1,1]$.

During training, we decay the learning rate after each batch of processing. The learning rate decay strategy conforms to exponential decay:
\begin{equation}
\eta=\eta\beta
\end{equation}
where $\eta$ denotes the learning rate and $\beta$ is the learning rate decay coefficient. The learning rate will decay until it reaches the preset minimum value. In order to avoid the instability during training, the gradients of the weight matrix are limited to the range of $[-1,1]$.

\subsection{Complexity Analysis}
The CNCA-IGE model necessitates the degree centrality and the sparse representation of the adjacency matrix as inputs, whic has a time complexity of $O(|V|)$ and $O(|E|)$, respectively. After that, the model performs a series of matrix multiplications, where these operations are bounded by the matrix operations using the adjacency matrix $A$, with dimension $V\times V$, resulting in a time complexity of $O(V^2)$. However, since we use sparse matrix representations, this complexity becomes associated with the density of the matrix, i.e, $O(|E|)$. Once the nodes have been encoded, we calculate their corresponding BC rankings, the time complexity of this process takes $O(|V|)$. Note that the CNCA-IGE computes the target centrality of all nodes in a single pass. As a result, the total complexity of the CNCA-IGE model is provided by $O\left(2\left|V\right|+\left(1+c_p\right)\left|E\right|\right)=O(|E|)$, where $c_p$ is a constant reflecting the number of operations performed by the proposed model. Notably, the training process is only done once here.

\begin{figure*}[htpb]
	\centering
	\includegraphics[width=0.9\linewidth]{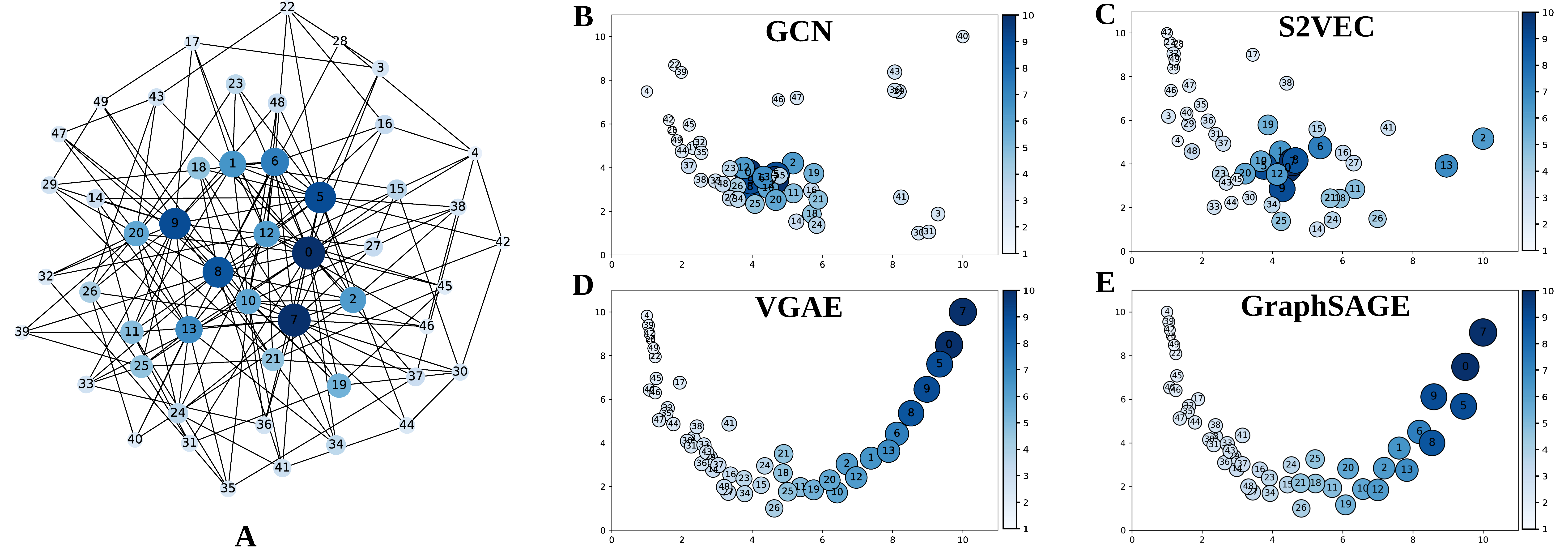}
	\caption{Visualizing Embeddings: A Case Study Using a synthetic network with 50 Nodes and an Average Degree of 7.36}
	\label{fig:07}
\end{figure*}

\section{Experiments}\label{sec4}

We utilize 2D Principal Component Analysis projection to visually represent our learned embeddings, providing an intuitive demonstration of our model's ability to preserve the relative CC order among nodes in the embedding space. For comparative analysis, we include results from two traditional node embedding models, namely GCN and S2VEC, to assess their capacity to maintain CC similarity. The example network is generated using the power-law cluster model (implemented with Networkx 2.6.3), with a specified number of nodes ($n = 50$) and an average degree ($m = 7.36$). All embedding dimensions are fixed at 256. As depicted in Fig. \ref{fig:07}, only in cases D and E do linearly separable segments correspond to clusters of nodes with similar CC, while the other two models fail to exhibit this pattern. This observation underscores the potential of inductive graph embedding methods to generate more discriminative embeddings for CC ranking prediction, which is a key factor contributing to their prediction accuracy in subsequent experiments.

\begin{algorithm}
	\caption{Train algorithm for CNCA-IGE}\label{alg1}
	\KwIn{Encoder parameters $W_{\text{en}}$ \\ \qquad \quad Decoder parameters $W_{\text{de}}$}
	\KwOut{Trained Model model.ckpt}
%	\tcp{Data Processing}s
	{Generate Small-World Networks $G_{\text{WS}}$ and Scale-Free Networks $G_{\text{BA}}$, $G = (G_{\text{WS}} \cup G_{\text{BA}})$} \\
	\For {$j \leftarrow 1$ \KwTo $M$}{ 
		\For {$i \leftarrow 1$ \KwTo $N$}{ 
			Calculate each node's $v_i$ degree centrality $ DC_i$ 
		}
            Compute each node's $v_i$ closeness centrality $CC_i$ and betweenness centrality $BC_i$, $\forall v_i \in G$ 
	}
%	\tcp{Train}
	\For{each epoch}{
		\For{$i \leftarrow 1$ \KwTo $N$}{ 
			Learn the graph embeddings $E_i$ using the Inductive Graph Encoder with       Eq. \eqref{4} and \eqref{5}\\
			\If{predicted centrality = Closeness centrality}{
				Embedding Decoder = MLP 
				
				 Compute $CC_i$ ranking with Eq. \eqref{9}
			}
			\ElseIf{predicted centrality = Betweeness centrality}{
				Embedding Decoder = MLP-Mixer
				
				Compute $BC_i$ ranking with Eq. \eqref{10}
			}
			Update $W_{\text{en}}$ and $W_{\text{de}}$ with Adam optimizer by minimizing Eq. \eqref{11}
		}
	}
\end{algorithm}

\begin{table*}[htbp]
	\centering
	\caption{Real World Complex Networks}
	\tabcolsep10.5pt
	\begin{tabular}{llccccc}
		\toprule
		Real-world Networks&{Abbreviation}& {Nodes} &{Edges} &  {Density} &Avg. Clustering Coef.& Avg. Degree\\
		\midrule
		email-Eu-core & Email & 1005  & 25571&   0.025 & 0.473 &33.246 \\
		p2p-Gnutella08 &   P2P-08 & 6301  & 20777 &  0.0005 &   0.015 &6.595 \\
		Erdos02.edges      & Erdos & 6927  & 11850  &  0.0002  &  0.398 &3.421   \\
		Lastfm\_asia\_edges    & LastFM & 7624  & 27806&    0.0004 &   0.285 &7.293       \\
		p2p-Gnutella09   & P2P-09 & 8114  & 26013 &   0.0004&    0.014& 6.412 \\
		p2p-Gnutella05 &     P2P-05 & 8846  & 31839 &   0.0004   & 0.009 &7.199  \\
		\bottomrule
	\end{tabular}%
	\label{tab:01}%
\end{table*}%

\subsection{Experimental Setup}
\subsubsection{Datasets}~\\
$~$\quad\textbf{Synthetic Networks:} Two sets of synthetic network data- sets: [i] scale free networks \cite{ba}; [ii] small world networks \cite{ws} are generated based on Barabási-Albert model and Watts-Strogatz model by the complex network generator in NetworkX as training sets. The Barabási-Albert model is used to generate scale free networks with a lognormal degree distribution, while the Watts-Strogatz model is used to generate small world networks with a degree distribution similar to a random graph. Both the Barabási-Albert model and the Watts-Strogatz model require two parameters, the number of nodes in the network and the number of edges connecting the new nodes to the established nodes during the network generation process. In order to approximate the real network, our training set contains a total of 600 synthetic networks, each with ranging from 100 to 1,000 nodes, making it small in size while having distributional properties that match those of the real network, facilitating the subsequent computation of the actual closeness centrality and betweenness centrality.

\textbf{Real World Networks:} The real-world networks are taken from the Stanford Large Network Dataset Collection, and the networks used in the test set and their properties are listed in the Table \ref{tab:01}. The density property reflects the concentration of connections within its adjacency matrix, the average clustering coefficient quantifies the extent to which nodes in a graph tend to cluster together, and the average degree refers to the average number of edges connected to each node within the network. 

\subsubsection{Baseline and Other Settings}~\\
$~$\quad\ For the closeness centrality sorting and betweenness centrality sorting prediction tasks, we select the GCN+MLP and S2VEC+MLP combinations as the baseline to compare with the methods proposed in this paper. For the baseline method, we perseverate the best results based on the parameter settings on the source code provided by the authors of the NCA-GE model. We implemented our approach using the TensorFlow deep learning framework and conducted model training on a compute server running Ubuntu 20.04, which was equipped with four NVIDIA GTX 1660 graphics processors.

\subsubsection{Evaluation Metrics}~\\
$~$\quad\ For all baseline methods and CNCA-IGE, we report their effectiveness in terms of kendall tau distance, and their efficiency in terms of wall-clock running time.

\textbf{Kendall tau-b} is a metric that quantifies the number of disagreements between compared methods’ rankings. The Kendall tau-b correlation coefficient is computed as follows:
\begin{equation}
K\left(\tau_1,\tau_2\right)=\frac{2(\alpha-\beta)}{n(n-1)}
\end{equation}

Where $\alpha$ is the number of concordant pairs, and $\beta$ is the number of discordant pairs. The value of kendall tall distance is in the range [-1, 1], where ``1" means that two rankings are in total agreement, while ``-1" means that the two rankings are in complete disagreement.

\textbf{Wall-clock running time} refers to the actual time taken from the start to the end of a computer program’s execution, typically measured in seconds.

\subsection{Performance and Discussion}
In our experimental setup, our primary focus is on the prediction task involving the ranking of centrality metrics. For each synthetic network dataset, the experiment divides it into training and test sets in the ratio of 8:2. Based on the existing synthetic network training set, synthetic network test set, and real-world complex network datasets, we set up both transductive and inductive task scenarios. In the transductive task scenario, the trained regression model is used to predict the closeness centrality ranking and the betweenness centrality ranking of the nodes in the synthetic network test set with the same degree distribution and clustering coefficients as the training set, while in the inductive task scenario, the regression model is used to predict the closeness centrality ranking and the betweenness centrality ranking of the nodes in the real-world complex network. This task highlights the model's adaptability to new, complex network structures, underscoring its generalization strengths. To ensure the robustness of our findings, all reported results are based on the average of 10 independent runs.
%
% \begin{table}[htbp]
% 	\centering
% 	\caption{Kendall tau-b for closeness centrality ranking}
% 	\tabcolsep2.3pt
% 	\begin{tabular}{@{}cccccccc@{}}
% 		\toprule
% 	  & Small world & P2P-05 & P2P-08 & P2P-09 & Erdos & LastFM & Email  \\
% 		\midrule
% 		GCN+MLP & 0.944       & 0.832 & 0.82  & 0.811 & 0.717 & 0.7   & 0.819   \\
% 		S2VEC+MLP & 0.952     & 0.827 & 0.813 & 0.801 & 0.683 & 0.698 & 0.833   \\
% 		VGAE+MLP & 0.967     & \textbf{0.933} & \textbf{0.92} & \textbf{0.912} & \textbf{0.79} & \textbf{0.718} & \textbf{0.841}  \\
% 		&       &       &       &       &       &       &         \\
%   & Scale free        & P2P-05 & P2P-08 & P2P-09 & Erdos & LastFM & Email \\
% 		GCN+MLP & 0.912     & 0.683 & 0.671 & 0.653 & 0.786 & 0.61  & 0.892 \\
% 		S2VEC+MLP & 0.94        & 0.677 & 0.64  & 0.649 & 0.804 & 0.637 & 0.897  \\
% 		VGAE+MLP & 0.961     & \textbf{0.834} & \textbf{0.811} & \textbf{0.797} & \textbf{0.874} & \textbf{0.722} & \textbf{0.924}   \\
% 		\bottomrule
% 	\end{tabular}%
% 	\label{tab:03}%
% \end{table}%

\begin{table*}[htbp]
    \centering
    \caption{Kendall Tau-b Correlation in Evaluating Closeness Centrality Rankings}
    \tabcolsep5pt
    \begin{tabular}{lcccccccccccccc}
    \toprule
     & \multicolumn{2}{c}{Train\&Test} & \multicolumn{2}{c}{P2P-05} & \multicolumn{2}{c}{P2P-08} & \multicolumn{2}{c}{P2P-09} & \multicolumn{2}{c}{Erdos} & \multicolumn{2}{c}{LastFM} & \multicolumn{2}{c}{Email} \\
    & WS & BA & WS & BA & WS & BA & WS & BA & WS & BA & WS & BA & WS & BA \\
    \midrule
    GCN+MLP & 0.944 & 0.912 & 0.832 & 0.683 & 0.82 & 0.671 & 0.811 & 0.653 & 0.717 & 0.786 & 0.7 & 0.61 & 0.819 & 0.892 \\
    S2VEC+MLP & 0.952 & 0.94 & 0.827 & 0.677 & 0.813 & 0.64 & 0.801 & 0.649 & 0.683 & 0.804 & 0.698 & 0.637 & 0.833 & 0.897 \\
    VGAE+MLP & 0.967 & 0.961 & \textbf{0.933} & \textbf{0.834} & \textbf{0.92} & \textbf{0.811} & \textbf{0.912} & \textbf{0.797} & \textbf{0.79} & \textbf{0.874} & \textbf{0.718} & \textbf{0.722} & \textbf{0.841} & \textbf{0.924} \\
    \bottomrule
    \end{tabular}
    \label{tab:03}
\end{table*}%

Table \ref{tab:03} provide a comprehensive overview of the model performances achieved through training on diverse network datasets for the prediction of closeness centrality rankings. The VGAE combined with MLP outperforms the baseline GCN/S2VEC with MLP model in both task scenarios across all datasets. The baseline model performs well in predicting the closeness centrality ranking of synthetically generated networks because these networks have similar degree distributions and clustering coefficients as the training set, However, the closeness centrality ranking prediction performance of the baseline model drops significantly when predicting the centrality ranking of real-world complex networks. This decline underscores the model's limitations in adapting to the intricacies of real-world network data. In contrast, VGAE, as an inductive graph embedding algorithm capable of generalizing patterns from existing data for application to new unknown data, has a clear advantage in generalization performance. The trained regression model still has satisfactory predictive performance in predicting the closeness centrality ranking of real-world complex networks. This clear advantage underscores VGAE's capacity to adapt to novel data with diverse characteristics and complexities.
\begin{table*}[ht]
    \centering
    \caption{Kendall Tau-b Correlation in Evaluating Betweenness Centrality Rankings}
    \tabcolsep5pt
    \begin{tabular}{lcccccccccccccc}
    \toprule
     & \multicolumn{2}{c}{Train\&Test} & \multicolumn{2}{c}{P2P-05} & \multicolumn{2}{c}{P2P-08} & \multicolumn{2}{c}{P2P-09} & \multicolumn{2}{c}{Erdos} & \multicolumn{2}{c}{LastFM} & \multicolumn{2}{c}{Email} \\
     % & \cline{2-3} & \cline{4-5} 
    & WS & BA & WS & BA & WS & BA & WS & BA & WS & BA & WS & BA & WS & BA \\
    \midrule
    GCN+MLP & 0.857 & 0.86 & 0.765 & 0.701 & 0.719 & 0.644 & 0.724 & 0.702 & 0.577 & 0.663 & 0.733 & 0.674 & 0.742 & 0.817 \\
    S2VEC+MLP & 0.871 & 0.863 & 0.779 & 0.733 & 0.709 & 0.693 & 0.709 & 0.708 & 0.532 & 0.597 & 0.717 & 0.661 & 0.723 & 0.77 \\
    VGAE+MLP & 0.884 & 0.872 & 0.844 & 0.79 & 0.82 & 0.801 & 0.813 & 0.793 & 0.719 & 0.741 & 0.795 & 0.75 & 0.801 & 0.831 \\
    GraphSAGE+MLP & 0.891 & 0.884 & \textbf{0.862} & \textbf{0.817} & \textbf{0.852} & \textbf{0.819} & \textbf{0.861} & \textbf{0.83} & \textbf{0.723} & \textbf{0.774} & \textbf{0.804} & \textbf{0.763} & \textbf{0.823} & \textbf{0.846} \\
    \bottomrule
    \end{tabular}
    \label{tab:04}
\end{table*}%

Table \ref{fig:04} details the performance analysis of models trained onvarious synthetic networks for predicting betweenness centrality rankings. In both the transductive task scenario and the inductive task scenario, our study employed the combined model of GraphSAGE/VGAE with MLP, which significantly outperformed the baseline model GCN/S2VEC with MLP on all real-world complex network datasets. Consistent with the results of the prediction experiments on closeness centrality ordering, there is a discernible decline in the performance of the baseline model when it comes to predicting centrality rankings for real-world complex networks, despite its commendable performance in estimating betweenness centrality ranking for synthetically generated networks. Conversely, the GraphSAGE and VGAE models, characterized by their inductive graph embedding capabilities, emerge as powerful contenders. Their distinct advantage lies in their robust generalization performance, which enables our trained regression model to continue delivering desirable prediction results when tasked with estimating betweenness centrality rankings for real-world complex networks.

It is worth noting that, when considered within the theoretical framework of complex networks, the P2P network can be approximated as a small-world network with exponential distribution of degree distribution, and the LastFM network is usually categorized as a social network. Therefore, P2P-05, P2P-08, P2P-09, and LastFM, as experimentally selected real-world complex networks, have network attribute characteristics more closely matching small-world networks. Conversely, Email and Erdos networks, recognized as classical email networks, exhibit more pronounced scale-free network properties in their network structures.  This classification is substantiated by our experimental results. Specifically, when we employ a small-world synthetic network as the training set, our model achieves superior centrality prediction performance for P2P-05, P2P-08, P2P-09, and LastFM networks compared to when using a scale-free synthetic network as the training set. In contrast, when confronted with Email and Erdos networks, the model's centrality prediction effect is not as robust as when trained on scale-free synthetic networks.
\begin{figure}[htpb]
	\centering
	\includegraphics[width=0.85\linewidth]{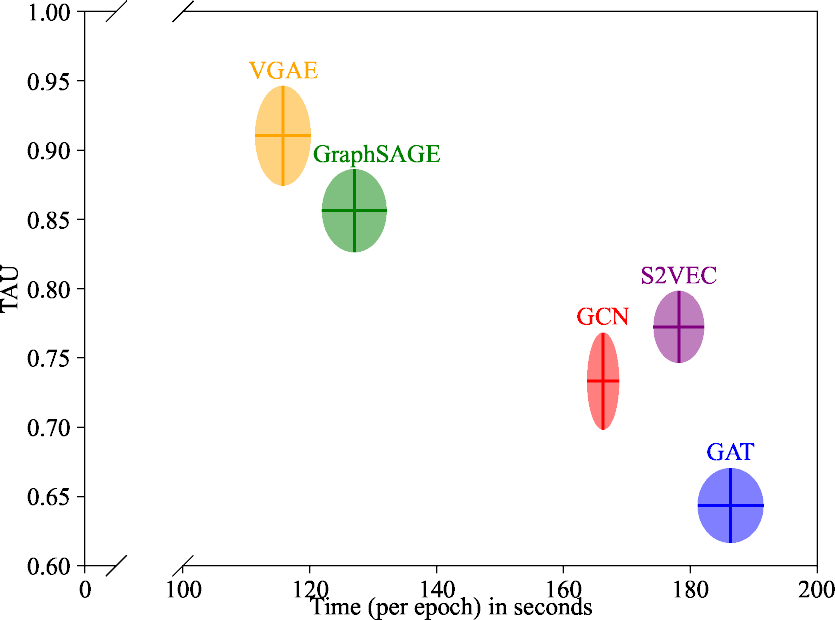}
	\caption{Performance-Speed Tradeoff: Comparing Kendall Tau-b and Model Speed}
	\label{fig:08}
\end{figure}

The GraphSAGE, VGAE, and MLP integrated model leverages parallel processing and sample aggregation to markedly cut training time for inductive tasks against the GCN/S2VEC and MLP model.  As shown in Fig. \ref{fig:08}, our proposed model has a 25\%-30\% reduction in training time compared to the baseline model while having considerably better centrality ranking prediction performance.

As shown in Fig. \ref{fig:01}, we notice that the correlation between betweenness centrality and degree centrality is more complex and nonlinear compared with that between closenss centrality and degree centrality. Empirical findings confirm that predicting betweenness centrality ranking is a more intricate task compared to closeness centrality ranking prediction, even when employing the same model architecture. Notably, the performance of the model, which combines GraphSAGE, VGAE, and MLP, exhibits a slightly lower accuracy in betweenness centrality ranking prediction compared to closeness centrality ranking prediction. Further more, the MLP-based decoder architecture is not stable enough as the embedding vector dimension fluctuates, resulting in a significant decline in the betweenness centrality ranking prediction performance as the embedding dimension increases. To solve these problems, in the betweenness centrality ranking prediction task, we introduce the MLP-Mixer model, which departs from the conventional MLP architecture by incorporating feature mixing within the node embedding vectors. In addition, MLP-Mixer employs neural network architecture components such as residual connectivity and layer normalization, which contribute to a more stable and robust prediction model.

\begin{figure}[htpb]
	\centering
	\includegraphics[width=0.85\linewidth]{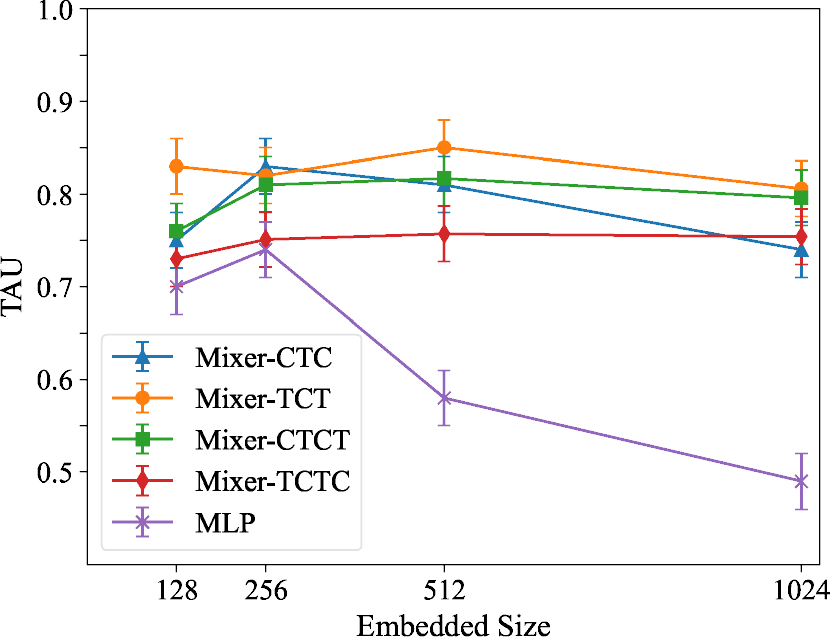}
	\caption{Performance Contrast: MLP-Mixer versus MLP at Various Embedding Dimensions}
	\label{fig:10}
\end{figure}

The MLP-Mixer employs two modules, token-mixing module and channel-mixing module, which correspond to the feature mixing within the node embedding vectors and the feature mixing between the nodes embedding, respectively. To assess the effectiveness of the MLP-Mixer neural network, we conduct a comparative analysis with the baseline model, MLP. To achieve optimal feature extraction, we investigate four distinct combinations of the token-mixing module and the channel-mixing module within the MLP-Mixer, specifically Channel-Token-Channel (CTC), Token-Channel-Token (TCT), Channel-Token-Channel-Token (CTCT), and Token-Channel-Token-Channel (TCTC). Subsequently, we evaluate the model's performance using embedding vectors of varying dimensions as input. As shown in Fig. \ref{fig:10}, the superiority of employing MLP-Mixer over the baseline model is readily apparent. MLP-Mixer excels because it comprehensively integrates the notion of feature mixing both within individual nodes and across nodes in the network. This approach, distinct from the baseline model, is characterized by a heightened model capacity, enabling it to capture and model complex nonlinear correlations more effectively. Furthermore, MLP-Mixer demonstrates superior stability as the embedding dimension varies. This stability, a key attribute, ensures consistent performance regardless of the dimensionality of the embedding vectors.

\section{Conclusion}\label{sec5}
In this paper, we propose a new architecture that combines the inductive graph embedding algorithms GraphSAGE and VGAE with MLP-Mixer neural network to train a regression model that approximate the closeness centrality ranking and betweenness centrality ranking of high computational complexity with low computational complexity degree centrality. Compared with existing methods which use transductive graph embedding method combined with MLP to train regression models, the architecture in this paper is optimized in terms of generalization performance and model capacity. Experimental results implies that our model outperforms most existing algorithms in closeness centrality and betweenness centrality ranking prediction scenarios for large-scale real-world networks. Meanwhile, compared to state-of-the-art baseline model, our model has 25\%-30\% reduction in training time, which makes it more suitable for dealing with the task of centrality metrics ranking prediction for large-scale real networks. 

Notably, the above results also highlight the importance of complex network models (e.g. small-world networks and scale-free networks). They capture critical features of real-world networks, which is often extremely important for training deep learning models to solve challenging problems in complex real-world networks. In future research, we aim to enhance CNCA-IGE's capacity for directed graphs with weighted edges using inductive graph embedding methods. Furthermore, investigating the potential of CNCA-IGE for temporal dynamic graphs can provide ideas for the study of time-varying networks and contribute to further understanding of the characteristics of temporal networks.

\bibliographystyle{IEEEtran}
\bibliography{IEEEabrv, reference.bib}

\begin{IEEEbiography}[{\includegraphics[width=1in,height=1.25in,clip,keepaspectratio]{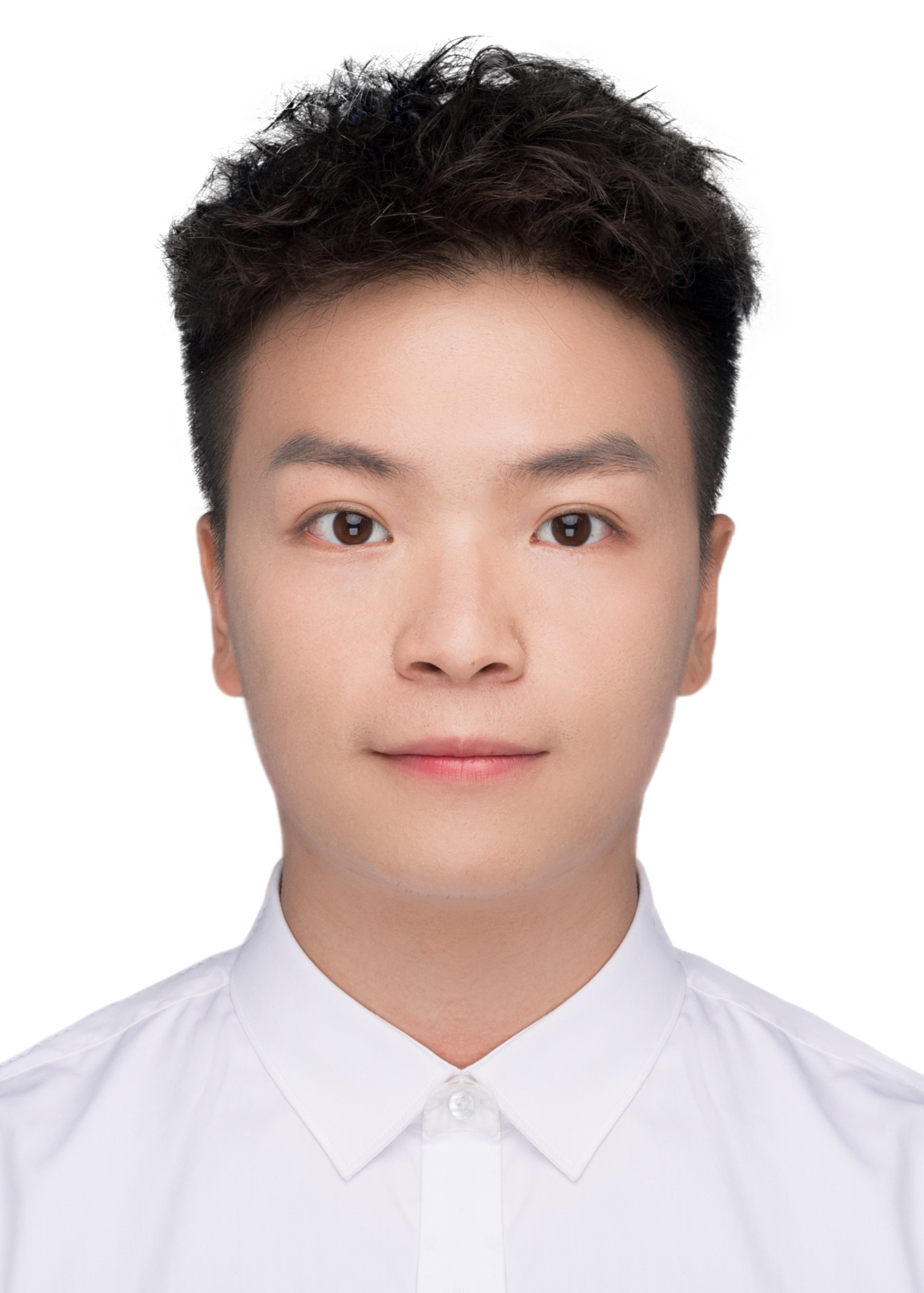}}]{Yiwei Zou}
graduated from Sun Yat-sen University with a bachelor's degree in Information Engineering and is now researching for a master's degree in Computer Science and Technology at Sun Yat-sen University. His main research field includes the exploration of graph neural network modeling and its application in the complex network.
\end{IEEEbiography}
\begin{IEEEbiography}[{\includegraphics[width=1in,height=1.25in,clip,keepaspectratio]{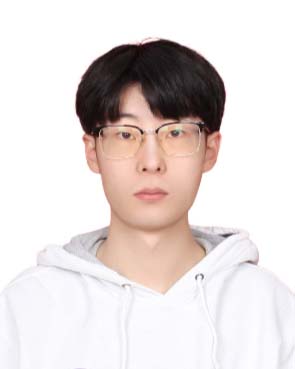}}]{Ting Li}
was born in Inner Mongolia, China, and received the B.Sc in Computer Science from Inner Mongolia University. He is currently working toward an M.Sc. degree in the School of Systems Science and Engineering at Sun Yat-sen University. His research is primarily focused on the application of machine learning for network science.
\end{IEEEbiography}
\begin{IEEEbiography}[{\includegraphics[width=1in,height=1.25in,clip,keepaspectratio]{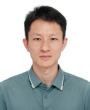}}]{Zong-Fu Luo}
received the B.E.Sc., M.E.Sc. and Ph.D. degrees in aeronautical and astronautical science and technology from National University of Defense University, Changsha, China, in 2007, 2010 and 2015, respectively. He was a visiting student with Politecnico di Milano between October 2012 and June 2014. He is currently an Associate Professor with Sun Yat-sen University, Guangzhou, China. Prior to joining SYSU, he was with Nanjing University as an Associate researcher between November 2018 and December 2020. His current research interests include complex network dynamics and intelligence Control.  

\end{IEEEbiography}

% 附录
\onecolumn
\appendix
\section{Appendix}

The training hyperparameter settings for all models are shown in Table \ref{tab:02}. 

\begin{table*}[htbp]
	\centering
	\caption{Hyperparameter Settings}
	\begin{tabular}{lllccccc}
		\toprule
		& Upstream & Downstream & Learning rate & Learning rate (min) & Embeded size & Batch size & $\lambda$ \\
		\midrule
		& VGAE  & MLP   & 0.001 & 0.000001 & 32    & 256   & 0.1 \\
		CLOSSNESS & GCN   & MLP   & 0.001 & 0.0001 & 128   & 1024  & 0.01 \\
		& S2VEC & MLP   & 0.001 & 0.0001 & 128   & 1024  & 0.1 \\
            \noalign{\vskip 3pt}
		\cline{2-8}
            \noalign{\vskip 3pt}
		& GraphSAGE & MLP-Mixer & 0.0001 & 0.000001 & 128   & 1024  & 0.1 \\
		\multirow{2}*{BETWEENNESS} & VGAE  & MLP-Mixer & 0.0001 & 0.000001 & 512   & 258   & 0.1 \\
		& GCN   & MLP   & 0.001 & 0.0001 & 128   & 1024  & 0.01 \\
		& S2VEC & MLP   & 0.001 & 0.0001 & 128   & 1024  & 0.1 \\
		\bottomrule
	\end{tabular}%
	\label{tab:02}%
\end{table*}%

\begin{figure*}[htpb]
	\centering
	\includegraphics[width=0.8\linewidth]{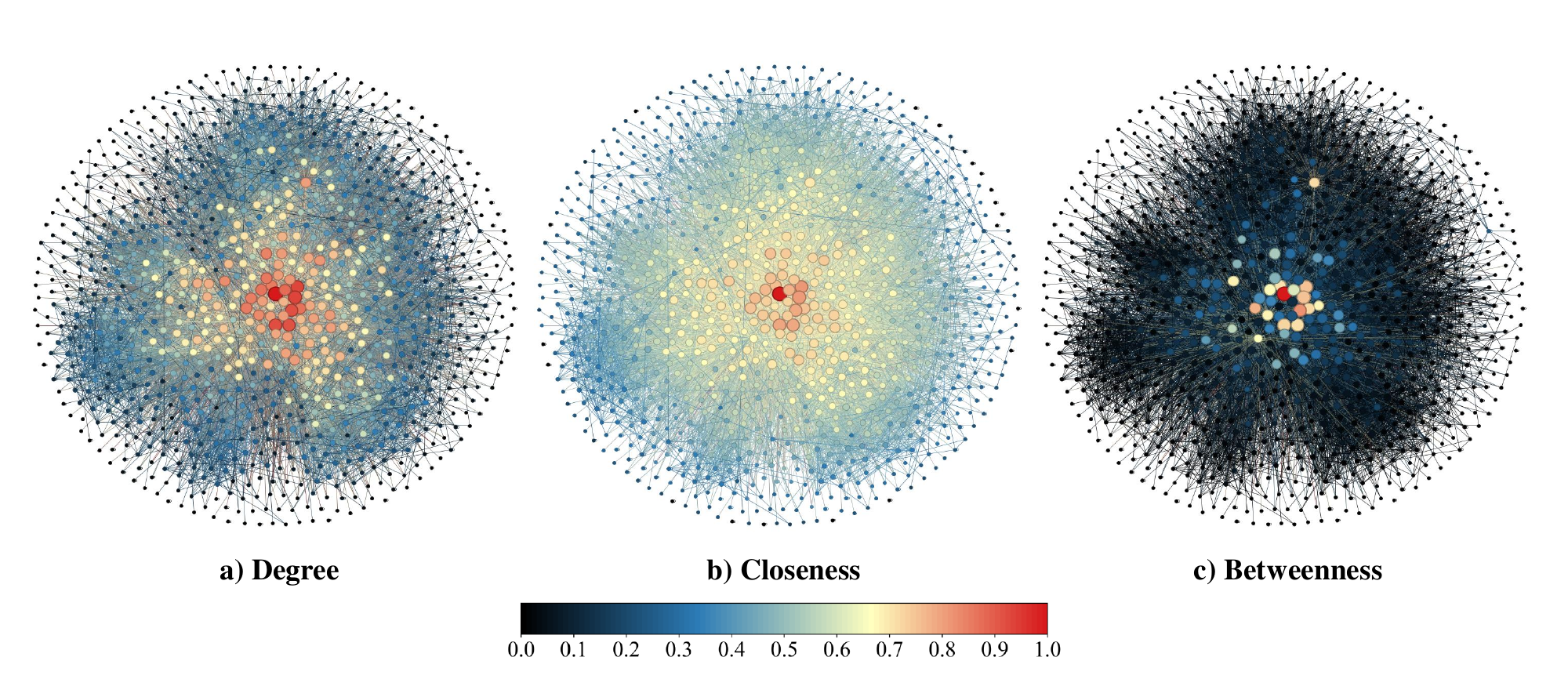}
	\caption{Centrality-Driven Network Visualization of email-Eu-core Networks}
	\label{fig:email-Eu-core}
\end{figure*}

\begin{figure*}[htpb]
	\centering
	\includegraphics[width=0.89\linewidth]{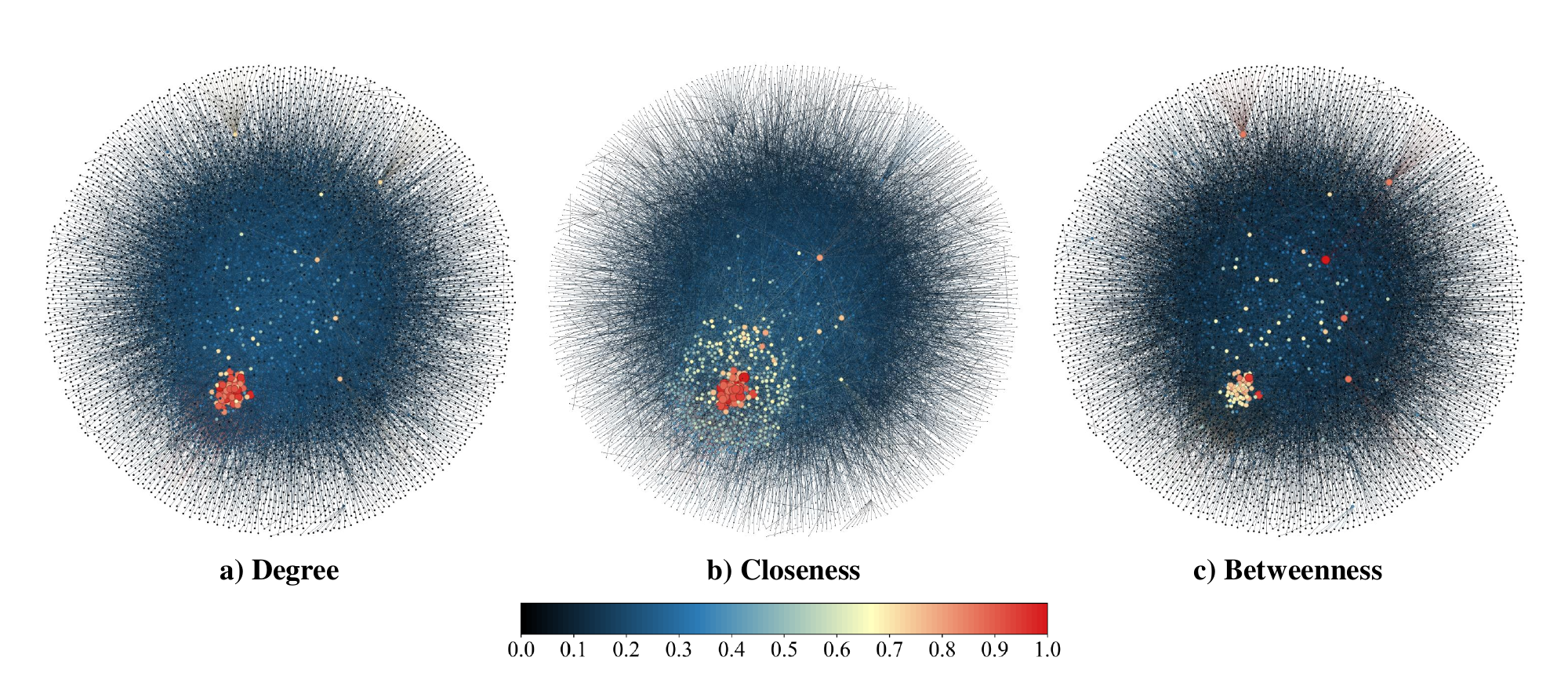}
	\caption{Centrality-Driven Visualisation of p2p-Gnutella08 Networks}
	\label{fig:appendix_02}
\end{figure*}

\begin{figure*}[htpb]
	\centering
	\includegraphics[width=0.89\linewidth]{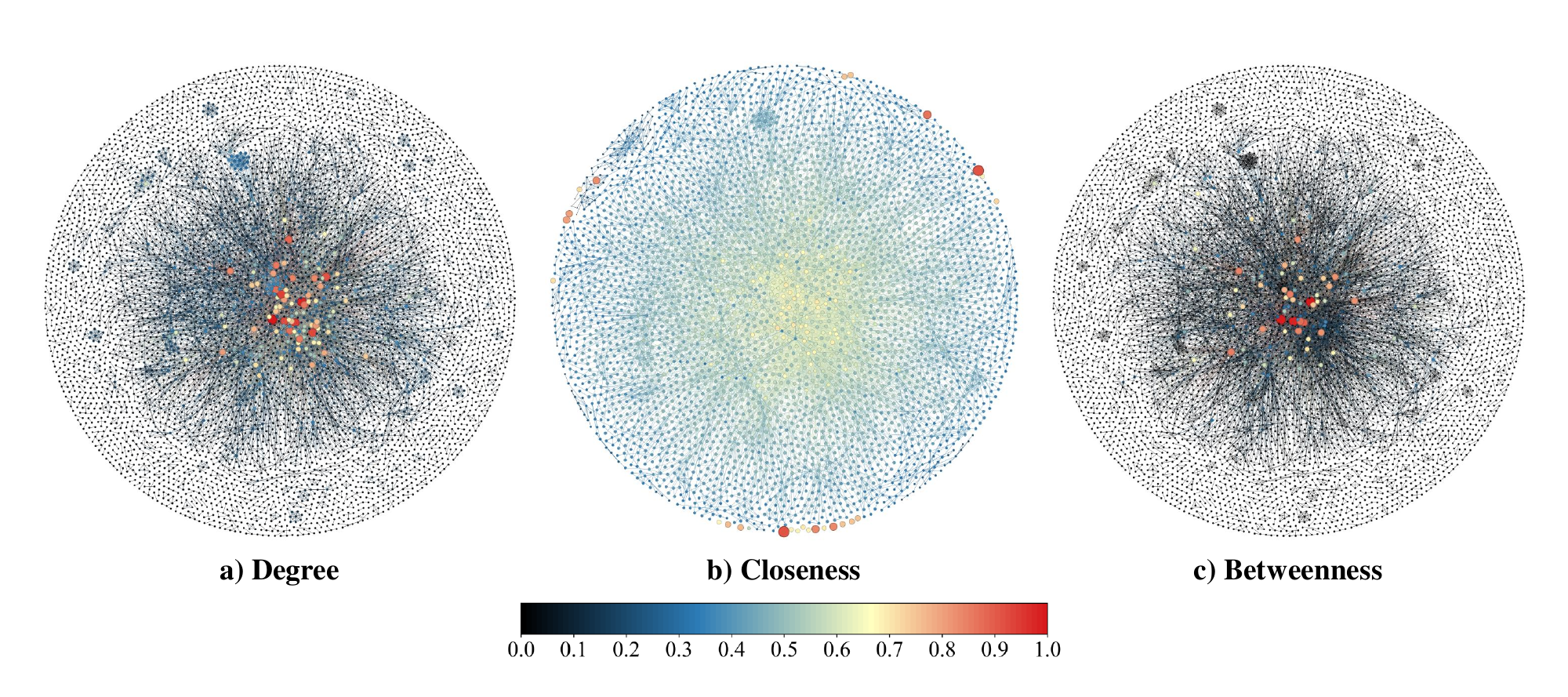}
	\caption{Centrality-Driven Visualisation of Erdos02.edges Networks}
	\label{fig:appendix_03}
\end{figure*}

\begin{figure*}[htpb]
	\centering
	\includegraphics[width=0.89\linewidth]{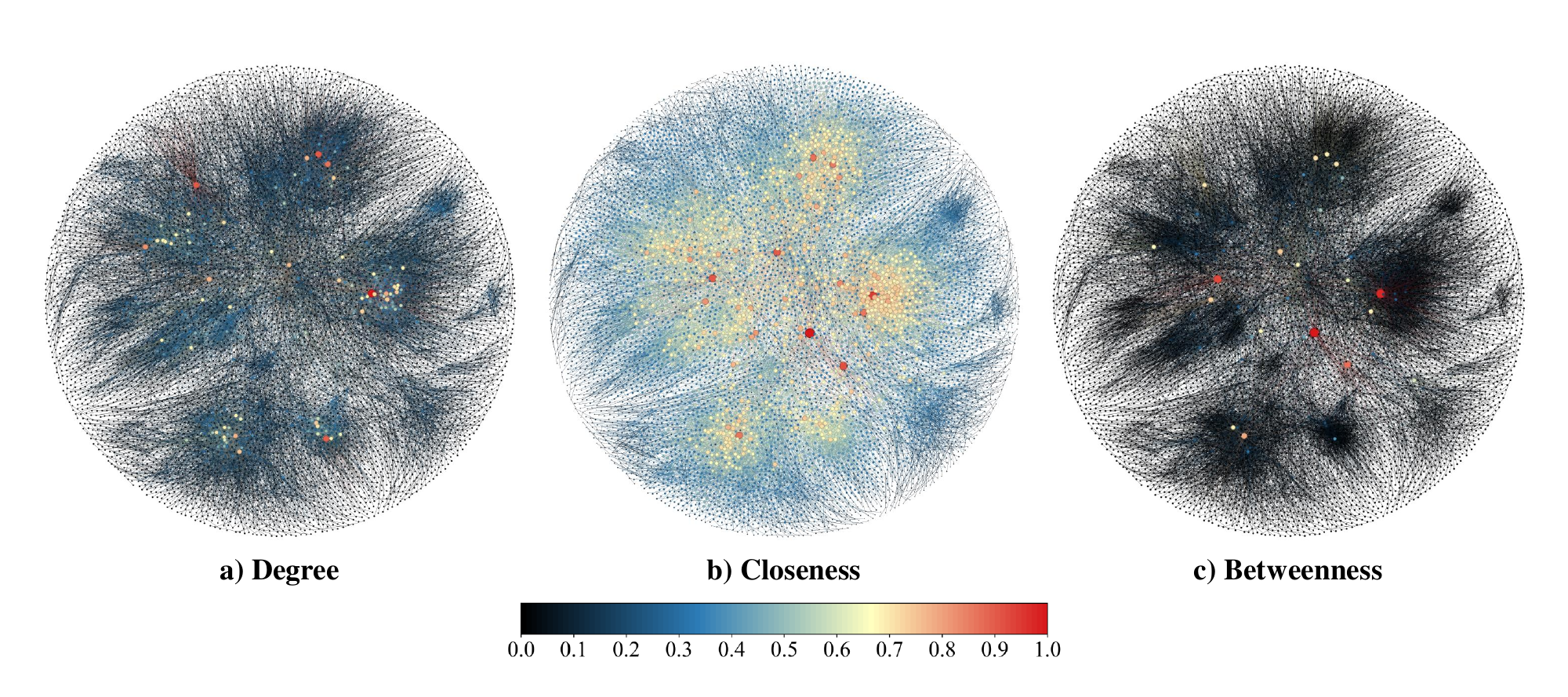}
	\caption{Centrality-Driven Visualisation of Lastfm\_asia\_edges Networks}
	\label{fig:appendix_04}
\end{figure*}

\begin{figure*}[htpb]
	\centering
	\includegraphics[width=0.89\linewidth]{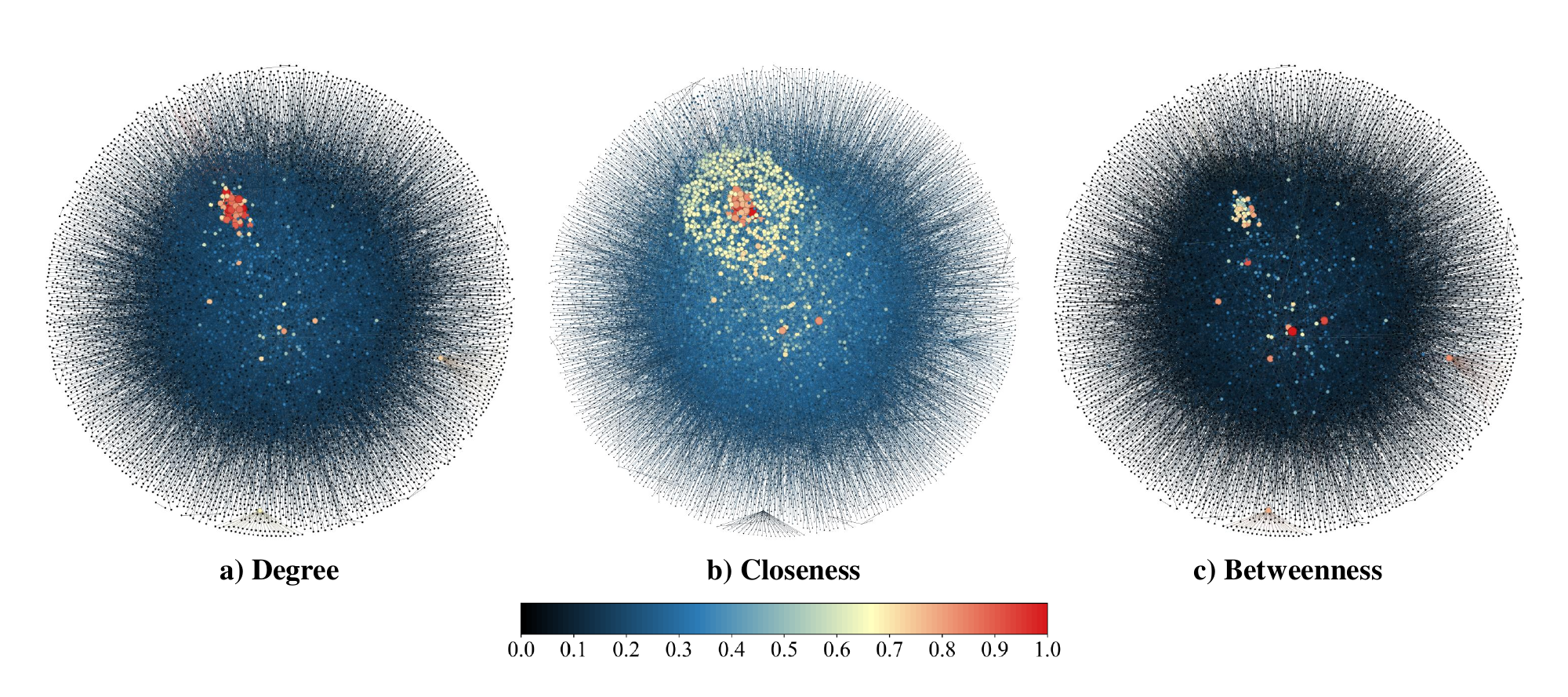}
	\caption{Centrality-Driven Visualisation of p2p-Gnutella09 Networks}
	\label{fig:appendix_05}
\end{figure*}

\begin{figure*}[htpb]
	\centering
	\includegraphics[width=0.89\linewidth]{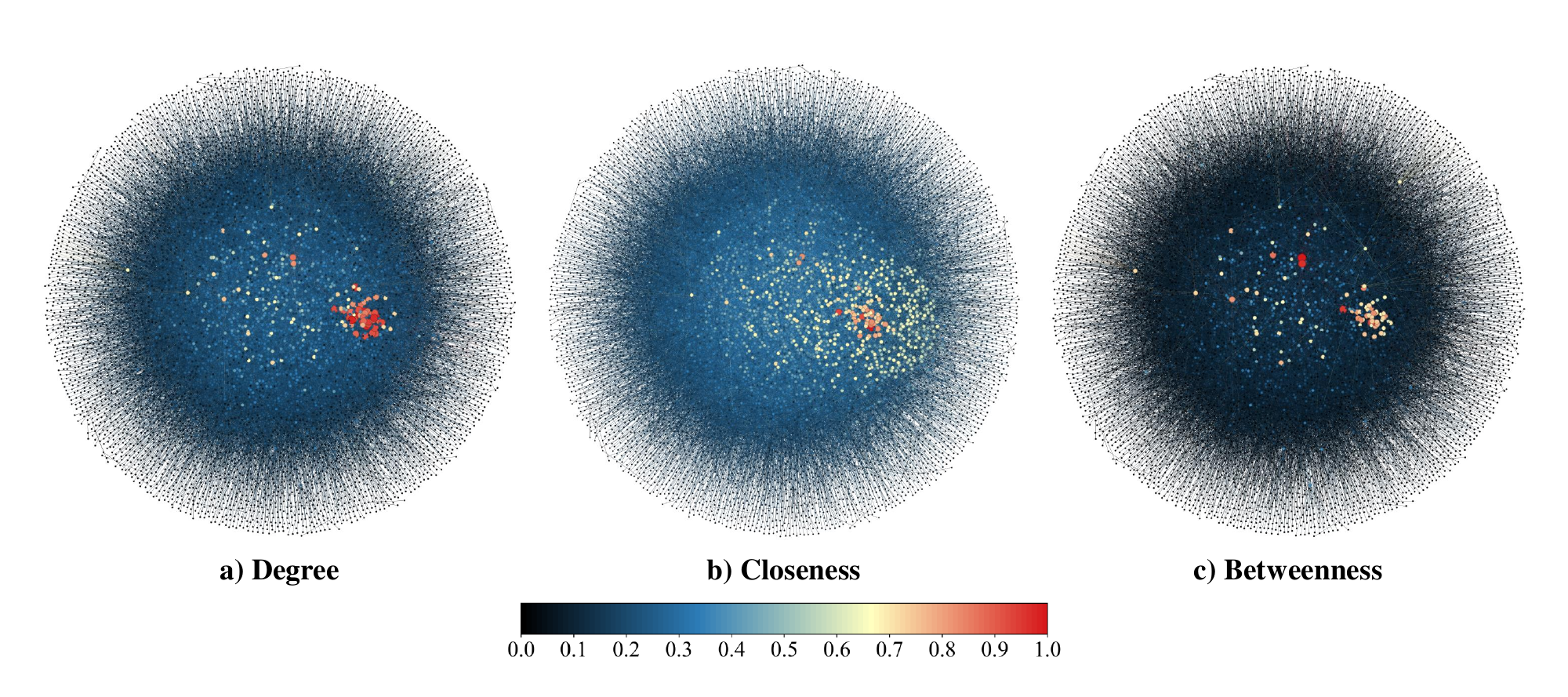}
	\caption{Centrality-Driven Visualisation of p2p-Gnutella05 Networks}
	\label{fig:appendix_06}
\end{figure*}

To present intuitively the correlation between the respective degree centrality, closeness centrality, and betweenness of the real-world complex networks selected in this paper, we visualise these networks as shown in 
Fig. \ref{fig:email-Eu-core} (email-Eu-core Networks), 
Fig. \ref{fig:appendix_02} (p2p-Gnutella08 Networks), Fig. \ref{fig:appendix_03} (Erdos02.edges Networks), Fig. \ref{fig:appendix_04} (Lastfm\_asia\_edges Networks), Fig. \ref{fig:appendix_05} (p2p-Gnutella09 Networks), and Fig. \ref{fig:appendix_06} (p2p-Gnutella05 Networks). Generally, degree centrality shows a significant correlation with closeness centrality and betweenness centrality, and nodes with relatively high degree centrality in the networks also tend to have much higher closeness centrality and betweenness centrality.

\end{document}